# C.V. Raman as a Science Communicator: A Historical Perspective

G.V. Pavan Kumar

Department of Physics, Indian Institute of Science Education and Research, Pune, India - 411008
email: pavan@iiserpune.ac.in
webpage: http://sites.iiserpune.ac.in/~pavan/

*C.V. Raman (1888 - 1970) was a creative scientist, enthusiastic teacher and a science celebrity in India. In all these roles, he communicated science effectively. In this essay, I ask how and why did he communicate science. I take a few examples from his research writings and show his ability to explain science lucidly. By looking into his thoughts on teaching and those of his students, I explore Raman, the teacher. Finally, I discuss a few aspects of his methods to communicate science to the public. I emphasize Raman's exposition and reveal a dichotomy.*

On 28th February 1928, K.S. Krishnan, the student, and C.V. Raman, his mentor, made a groundbreaking observation of a new type of radiation[1,2] from the light scattered by molecules in a liquid state. This observation shifted a paradigm in light-matter interaction from the so-called 'feeble fluorescence' to 'modified scattering'. What was considered a strange observation became an important result in quantum physics. Thus, the 'Raman effect' was born, and a new chapter was created in the history of science[3]. This was a discovery that involved creativity, perseverance, and collective effort led by C.V. Raman.

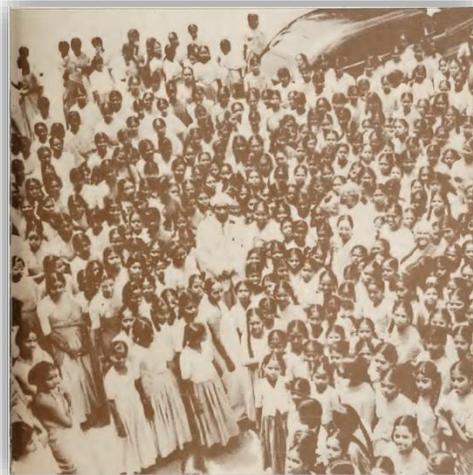

*Figure 1 Raman with school children. On specific occasions, he invited them to his institute and gave engaging science talks. Reproduced from his pictorial biography (see ref. 4, available on public archives)*

Every year, on 28th February, India celebrates National Science Day to commemorate the observation of the 'Raman effect'. Over the past few years, on this day, various institutions in India, including schools, colleges, and research institutes, interact with the public by communicating interesting and important aspects of science and technology. Some scientifically oriented industries too have joined this celebration. A great scientific discovery from India has created a nationwide carnival of science. At the heart of this carnival is the communication of science and an outreach to make science more accessible to public. Raman himself did this very efficiently.

With this background, it is relevant to revisit the life and times of Raman in the context of how he communicated science to various audiences. As a scientist, he played three different roles. First, he was a researcher working on cutting-edge research problems and communicated his results to the world, mainly through research journals and scientific talks. Second, Raman played an active role as a teacher. He trained a band of outstanding scientists (mostly men) and did this effectively by communicating the breadth and depth of his research field. Third, he communicated science to the general public (see Figure 1). He did this by giving talks in person, and by using the available mass media platforms, such as newspaper articles and radio broadcasts. In this essay, I discuss how and why Raman communicated science in all these roles.

## Raman as a researcher

As scientists, it is important to communicate your thoughts clearly to your peers. This communication happens mainly in journal publications, scientific talks, conference proceedings, and through patents. Raman was a very active researcher who constantly communicated his work through journal publications. During his career, he published more than 400 research papers[4,5], many of which changed how we view the physical world. His work was mainly related to experimental physics, specifically optics and spectroscopy. Natural and everyday phenomena significantly inspired his research questions.

Experimental physics connects abstract theoretical ideas and questions with the real world. This connection is made via careful measurement in a systematic and controlled way. Once the measurement is confirmed, academically-oriented scientists communicate their results in the form of a research paper that is subjected to peer review. Raman was a creative experimentalist. He designed and developed various methods to test and verify theoretical ideas, which needed a good grasp of the existing knowledge. Importantly, he effectively communicated his research results in plain and simple English. Now, I will discuss his most famous paper, that eventually led to the Nobel prize.

Since early 1920s, Raman and his research group were studying, light-matter interaction in the context of light scattering. Eventually, this study culminated into an effect, which was later to be called as the 'Raman effect'. The discovery of the this novel effect was communicated in two papers. One was published in *Nature*[2] and the other one was in *Indian Journal of Physics*[1]. The former paper was a crisp and quick description of the observed effect, and the latter was an oral presentation given to the South Indian Science Association on 16th March,1928 in Bangalore.

For this discussion, I would like to highlight some features of the paper published in *Indian Journal of Physics* as it contains more details and shows Raman's ability to communicate effectively. In this paper, Raman elaborated the history of the discovery of the 'new radiation'. He starts with various origins of radiation emitted by atoms and molecules. Among them, he identifies light scattering as one of the important processes. He states the difference between the primary (no energy change compared to incident light) and secondary radiation (with change in energy compared to incident light) that is emitted from atoms and molecules and thus introduces the reader to some of the fundamental aspects of light-matter interaction. Then he gives an overview of the work done in past 7 years or so in Calcutta (see[6] for more on Raman in Calcutta). He mentions about the optics experiments that he and his students were exploring. Specifically, he identifies the work done with 4 students - Seshagiri Rao, K.R. Ramanathan, KS Krishnan and S. Venkateshwaran. Raman talks about the roadblocks they had to face and why they had to temporarily stop some of the experiments. The main reason being fluorescent impurities in their experimental samples. This motivated them to initially call the optical phenomenon 'feeble fluorescence', which was essentially an extension of a known phenomenon. In the paper, Raman further reveals a crucial twist in the exploration. This was when he made a connection between his group's observation and the recently discovered X-ray scattering by Compton[7]. This motivated Raman to relook at all the observations that he and his group had been making over the past few years. He postulated that a possible optical analog of the Compton effect is the reason behind the new observations in his lab. This realization was a turning point in the discovery. At the same time, Raman and his group needed to ensure that the observed effect was indeed a new phenomenon and not the conventional fluorescence. Raman elaborates on this aspect as follows[1]:

> "You may therefore rightly ask me the question how does this phenomenon differ from fluorescence? The answer to the question is, firstly, that it is of an entirely different order of intensity. A more satisfactory proof was however forthcoming when Mr Krishnan and myself examined the polarisation of this new type of radiation and found that it was nearly as strong as that of the ordinary light scattering in many cases, and is thus quite 'distinct from ordinary fluorescence which is usually unpolarised."

This *polarised* nature of the scattered light is what makes the observed scattering different from anything that was observed until then. I need to emphasize that all these observations were done using sunlight along with a combination of colour filters. This was further verified with monochromatic lamps of the day, that were chosen and optimized for the experiments. I need to remind the readers that there were no lasers then (1920s).

Another important aspect discussed in the paper is the connection to quantum theory, in which Raman uses the concept - '*quantum of radiation*' and uses it to explain the observed signatures of light. In this context, Raman makes a connection to Kramers-Heisenberg's theory of dispersion[8] which was based on second-order perturbation theory in quantum mechanics and explains the possible transition in molecules due to illumination of light. He further makes some important comments on the connection between his observations with thermodynamics and coherence (or the lack of) in the scattered light.

The whole manuscript uses simple language, and communicates the ideas in an effective way. It is worth noting that he ends his paper with three things: an optimistic note, an acknowledgment, and an explicit mention of the date of this discovery[1]:

> "We are obviously only at the fringe of a fascinating new region of experimental research which promises to throw light on diverse problems relating to radiation and wave-theory, X-ray optics, atomic and molecular spectra, fluorescence and scattering, thermodynamics and chemistry. It all remains to be worked out. I have to add in conclusion that I owe much to the valuable co-operation in this research

> of Mr K S Krishnan, and the assistance of Mr S Venkateswaran and other workers in my laboratory. The line spectrum of the new radiation was first seen on the 28th February 1928. The observation was given publicity the following day."

Perhaps, this was the first *scientific presentation* in which Raman identifies the date of the important observation. Later, KS Krishnan's diary reconfirmed and elaborated the gripping story of this great discovery[5,9].

For this work, C.V. Raman was awarded the 1930 Nobel Prize in Physics[10]. In his Nobel lecture, he gave and insight into his discovery, and described his famous journey over the Mediterranean, which motivated him to ask questions relevant to light scattering[11] :

> "In the history of science, we often find that the study of some natural phenomenon has been the starting-point in the development of a new branch of knowledge. We have an instance of this in the colour of skylight, which has inspired numerous optical investigations, and the explanation of which, proposed by the late Lord Rayleigh, and subsequently verified by observation, forms the beginning of our knowledge of the subject of this lecture. Even more striking, though not so familiar to all, is the colour exhibited by oceanic waters. A voyage to Europe in the summer of 1921 gave me the first opportunity of observing the wonderful blue opalescence of the Mediterranean Sea. It seemed not unlikely that the phenomenon owed its origin to the scattering of sunlight by the molecules of the water. To test this explanation, it appeared desirable to ascertain the laws governing the diffusion of light in liquids, and experiments with this object were started immediately on my return to Calcutta in September, 1921. It soon became evident, however, that the subject possessed a significance extending far beyond the special purpose for which the work was undertaken, and that it offered unlimited scope for research. …"

These words indicate Raman's clarity of thought and expression. With simple language, he describes the history of his discovery. This kind of elegant science communication was one of the hallmarks of Raman's research papers and talks. The exposition was elegant and always to the point. He was also conversant in German and could give scientific talks. For example, he spoke about crystal dynamics in one of the Lindau meetings in 1950s[12]. This talk is a good example of his scientific communication skills.

Raman did all his work in India. His career spanned across many decades, and he worked in colonial and independent India. During these time, he communicated extensively with researchers across the world, including scientists in Europe and the US[13]. With good communication skills, he exchanged his ideas with his peers, which further played an important role in his thinking and subsequent publication of results. To generate such results he had to train students. Next, I will discuss how Raman communicated science with his students, and what they had to say about him as a teacher.

## Raman as a teacher

Experimental research is a collective endeavour that needs many minds and hands. Training a young generation of students who can directly participate in the experimental measurements is necessary. Raman trained many students during his tenure at Calcutta and Bangalore. To bring them to the standards of cutting-edge research, Raman taught fundamental physics and experimental methods used in his lab. He believed that teachers exposed to research can bring an important perspective to science. Importantly, he emphasized that questioning attitude developed by researchers can positively affect teaching.

Writing in 1971, L.A. Ramadas [1], one of Raman's students, says (as quoted in [14]):

> "He held the view that when a leading research worker takes on some special teaching course, he brings to his teaching the freshness of research and the questioning attitude which makes all the difference between dull pedagogy and inspired teaching. To some of us who had joined the M.Sc. course at Calcutta (both myself and the late Dr. K.S. Krishnan had joined the M.Sc. course in physics at Calcutta by 1920) Prof. Raman once made the side remark that the best way for him to master or revise any subject in physics was indeed to lecture on it to the M.Sc. classes."

As you may observe, Raman saw the utility of teaching not only in educating students but also in positively affecting the teacher. Raman had a deep appreciation for classical and quantum physics, and ensured that his students got the flavour of these subjects. Among many topics that he discussed with the students, electricity, magnetism and physical optics fascinated his students. As Ramdas mentions (as quoted in [14]) :

> "In listening spell-bound to Prof. Raman's lectures in "Electricity and Magnetism", covering a series of nearly 30 lectures, we had shared with him much of the excitement and superb thrill that Benjamin Franklin, Oersted, Arago, Gauss, Faraday, Maxwell, Hertz, Lord Kelvin and many others must have felt while they were making their actual discoveries. This was indeed no routine text-book learning, but reliving the actual past history of the subject. Almost regularly, Prof. Raman, with his genius for the subject, his extraordinary eloquence, imagery and fullness of precise expression, used to forget himself as well as the time and used to lecture for far more than the prescribed one hour, while the next lecturer was politely (and perhaps with a sense of relief) retiring from the scene after seeing Prof. Raman still at his lecture! In "Physical Optics", a topic on which he himself was conducting several investigations at the time, the students were introduced to all the topics coming, as it were, hot from the 'Lab', and the lecturer's flair for dipping straight into the great masters like Huygens, Fresnel, Mascart, Schuster, Wood, Rayleigh, and others of the late 19th and the early 20th centuries, imbued the students with a real love and enthusiasm for what they learned at Prof. Raman's feet, as it were."

---

[1] known for the Ramdas effect, related to lifted temperature minima

What were some of the methods used by CV Raman, the teacher? How did he conduct a class? What elements of pedagogy did he use? These are some of the interesting questions to address. The following excerpt from Raman's talk gives an insight into his methods[15]:

> "A study of the history of individual branches of science and of the biographies of the leading contributors to their development is essential for a proper appreciation of the real meaning and spirit of science. They often afford much more stimulating reading than the most learned of formal treatises on science. To the teacher, such histories and biographies are invaluable. Whenever he finds the attention of his listeners flagging a little, he can always enliven his class by telling a little story of how this or that great discovery in his subject was made or by recalling some anecdote about one or another of the famous investigators in the field. In this way, the teacher can convey to the student an understanding of how science is made and of the intellectual outlook which is the essence of it."

It is evident that Raman was a captivating lecturer and showed a lot of enthusiasm as a teacher. But, was he an inclusive mentor ? This question is yet to be resolved. An interesting aspect to note is that most of his students were men. We do not have enough information about what his women students thought about his teaching and interaction. In recent years in India, the issue of gender equity in science has come to focus, and Raman has been criticized for not being inclusive enough[16,17,18]. Raman was a difficult person and had many administrative conflicts[19,20]. This has been acknowledged by many people who knew him and interacted with him[5,14]. Perhaps he was more comfortable discussing and doing science at an individual level, than interacting with people, especially in administrative matters. However, his ability and commitment to communicate science seems to high. Furthermore, he enjoyed talking about science with a general audience, which I will discuss next.

## Raman's communication with public

One of the hallmarks of Raman as a scientist was that he engaged with the public, especially with children (see Figure 1), and communicated the excitement of science to them. Although he had many tussles with administrators and fellow scientists, he did not hesitate to interact with the public and opened his laboratory doors for them to visit occasionally. He enjoyed this public communication part of the work, and the audience reciprocated with enthusiasm. This is mainly because Raman was an excellent and engaging orator, as his former student Ramdas describes (as reproduced in[14]):

> "Raman excelled in public speaking and could give a lecture on, for instance, Egyptian History, off the cuff. His scientific lectures were a treat, for he was a superb entertainer. They were delivered in a high pitched resonant voice which reached the entire audience, making loud-speakers unnecessary. Rich in imagery and eloquence, and replete with spontaneous jokes, the lectures were given in so popular a style that every listener felt that he understood all the science that the learned lecturer was discussing."

This ability to connect with the audience is not a common thing for a scientist. A criticism of scientists, even today, is that they do not communicate well with the general audience.

Raman was an exception. He could have easily sat on his laurels and stayed away from the public eye, but he did not hesitate to express his thoughts in the public domain. Of course, he had the stature to do so. Even then, this was unusual for a scientist of his time.

In 1938, Raman gave a radio talk, titled *The new physics: Revolutionary advances*[15]. In there, he spoke about emerging physics in that era and the scientists who had contributed to it. Below is an excerpt from his radio talk :

> "I will not fatigue you by a recital of the names of even the most outstanding investigators who have built up the physics of today. Their names and their discoveries are known to every student of physics. They are claimed as nationals by one or another of many different countries. Yet in the truest sense, they belong to the whole world and to the International Brotherhood of Science. I will permit myself to mention only two of the greatest pioneers: Amongst the priceless memories that a man of science like myself treasures life-long is that of personal contact with such leaders of science as the late Lord Rutherford and the late Madame Curie; their contributions to the building of the New Physics have been most impressive, and their influence on their generation and on the progress of science almost incredibly great."

Raman admired many of the pioneers of modern physics, including Rutherford and Curie. He deeply appreciated science as a collective human endeavor, which is evident from his speech. As you may observe, there is clarity of thought in these communications, and this further highlights Raman's ability to connect with a broader audience.

## Raman contained multitudes

J.D. Bernal, in one of his essays[21], talks about the reciprocity between science and society. Each can feed the other, and this interaction has many stakeholders. At the center of this interface are scientists. Raman was well aware of this central role. He believed that science could influence society, and many of his writings and interactions on public forums showed his intention. He felt that science could uplift humankind and had a dominant role in influencing the lives of Indians, and the world. This thought was partly driven by the fact that Raman himself was a subject of colonial times, and he saw India undergo a transition from a colonial state to an independent one. He strongly believed that science had an important role in building the nation.

Raman donned the hats of a researcher, teacher, and public figure. He enjoyed researching and communicating science to a varied audience. From the existing records, it is clear that he was a complex person containing multitudes. He did have differences of opinion and intellectual fights with many people. But what drove him was his science and its communication. It is fascinating to see how a person who fought with so many people was so good at communicating his thoughts as a scientist. There is an interesting dichotomy in his character that needs further analysis.

## References:

1.  C.V. Raman, "A new radiation," Indian J Phys **2**, 387–398 (1928).

2. C. V. Raman and K. S. Krishnan, "A New Type of Secondary Radiation," 3048, Nature **121**(3048), 501–502, Nature Publishing Group (1928) [doi:10.1038/121501c0].

3. R. S. Krishnan and R. K. Shankar, "Raman effect: History of the discovery," Journal of Raman Spectroscopy **10**(1), 1–8 (1981) [doi:10.1002/jrs.1250100103].

4. S. Ramaseshan and C. Ramachandra Rao, *C.V. Raman : A Pictorial Biography*, Indian Academy of Sciences (1988).

5. G. Venkataraman, *Journey into Light: Life and Science of C.V. Raman*, Indian Academy of Sciences, Bangalore (1989).

6. P. Mukherji and A. Mukhopadhyay, *History of the Calcutta School of Physical Sciences*, 1st ed. 2018 edition, Springer, New York, NY (2018).

7. A. H. Compton, "A Quantum Theory of the Scattering of X-rays by Light Elements," Phys. Rev. **21**(5), 483–502, American Physical Society (1923) [doi:10.1103/PhysRev.21.483].

8. H. A. Kramers and W. Heisenberg, "Über die Streuung von Strahlung durch Atome," Z. Physik **31**(1), 681–708 (1925) [doi:10.1007/BF02980624].

9. D. C. V. Mallik, "The Raman effect and Krishnan's diary," Notes and Records of the Royal Society of London **54**(1), 67–83, Royal Society (2000) [doi:10.1098/rsnr.2000.0097].

10. R. Singh and F. Riess, "Sir C. V. Raman and the story of the Nobel prize," Current Science **75**(9), 965–971, Temporary Publisher (1998).

11. "The Nobel Prize in Physics 1930," NobelPrize.org, <https://www.nobelprize.org/prizes/physics/1930/raman/lecture/> (accessed 23 February 2024).

12. "Sir Chandrasekhara Raman - Lectures | Lindau Mediatheque," Lindau Nobel Mediatheque, 25 July 2014, <https://mediatheque.lindau-nobel.org/videos/31549/physics-of-crystals-german-and-english-presentation-1956/meeting-1956> (accessed 23 February 2024).

13. R. Singh, "CV Raman and the American scientists," INDIAN JOURNAL OF HISTORY OF SCIENCE (2003).

14. A. Jayaraman, *C.V.Raman: A Memoir*, Affiliated East-West Press,India (1990).

15. "The new physics: Revolutionary advances - Sir C.V. Raman's radio talk" (1938).

16. A. Sur, *Dispersed Radiance: Caste, Gender, and Modern Science in India*, Navayana, New Delhi (2011).

17. S. Deepika, "Women and the Institute," in Connect with IISc, pp. 2–3 (2018).

18. N. Jayaraj and A. Freidog, *Lab Hopping: Women Scientists in India*, Penguin Viking (2023).

19. R. S. Anderson, "THREE. The Bangalore Affair, 1935–38: Scientists and Conflict around C. V. Raman," in Nucleus and Nation - Scientists, International Networks, and Power in India, pp. 57–78, University of Chicago Press (2010) [doi:10.7208/9780226019772-008].

20. J. Phalkey, *Atomic State: Big Science In Twentieth-century India*, Permanent Black (2019).

21. J. D. Bernal, "The Social Relations of Science," Journal of the Royal Society of Arts **93**(4697), 458–464, Royal Society for the Encouragement of Arts, Manufactures and Commerce (1945).